\documentclass[conference]{IEEEtran}
\IEEEoverridecommandlockouts
\usepackage{amsmath,amssymb,amsfonts}
\usepackage{algorithmic}
\usepackage{graphicx}
\usepackage{textcomp}
\usepackage{xcolor}
\def\BibTeX{{\rm B\kern-.05em{\sc i\kern-.025em b}\kern-.08em
    T\kern-.1667em\lower.7ex\hbox{E}\kern-.125emX}}

\usepackage[T1]{fontenc}
\usepackage[nolist]{acronym}
\usepackage[style=numeric,sorting=none]{biblatex} 
\usepackage{xurl} 
\usepackage{svg}
\usepackage{amsmath}
\usepackage{booktabs}

\begin{document}
\title{Fast-Response Balancing Capacity of Alkaline Electrolyzers\\}
\author{
\IEEEauthorblockN{1\textsuperscript{st} Marvin Dorn\textsuperscript{1}\\0009-0005-1885-7031}
\and
\IEEEauthorblockN{2\textsuperscript{nd} Julian Hoffmann\textsuperscript{1}\\0000-0002-4105-4123}
\and
\IEEEauthorblockN{3\textsuperscript{rd}André Weber\textsuperscript{2}\\0000-0003-1744-3732}
\and
\IEEEauthorblockN{4\textsuperscript{th}Veit Hagenmeyer\textsuperscript{1}\\0000-0002-3572-9083}
\and
\IEEEauthorblockA{\textsuperscript{1}\textit{Karlsruhe Institute of Technology, Institute for Automation and Applied Informatics (IAI), Karlsruhe, Germany}}
\IEEEauthorblockA{\textsuperscript{2}\textit{Karlsruhe Institute of Technology, I. for Applied Materials - Electrochemical Technologies (IAM-ET), Karlsruhe, Germany}}

}

\begin{acronym}
\acro{afrr}[aFRR]{automatic Frequency Restoration Reserve}
\acro{aem}[AEM]{Anion Exchange Membrane}
\acro{ael}[AEL]{Alkaline Electrolysis}
\acro{bess}[BESS]{battery energy storage systems}
\acro{capex}[CAPEX]{Capital expenditures}
\acrodef{dod}[DoD]{Depth of Discharge}
\acrodef{eol}[EoL]{End of Life}
\acro{fcr}[FCR]{Frequency Containment Reserve}
\acro{fec}[FEC]{Full Equivalent Cycles}
\acro{gt}[GT]{gas turbine}
\acro{h2reb}[H2REB]{HydrogREenBoost}
\acro{hto}[HTO]{Hydrogen to Oxygen}
\acro{lfp}[LFP]{lithium iron phosphate battery}
\acrodef{lto}[LTO]{lithium titanate oxide}
\acro{matlab}[MATLAB\textregistered]{MathWorks\textregistered-MATLAB}
\acro{mfrr}[mFRR]{manual Frequency Restoration Reserve}
\acro{pem}[PEM]{Proton Exchange Membrane}
\acro{pv}[PV]{photovoltaics}
\acro{res}[RES]{renewable energy sources}
\acrodef{soc}[SoC]{State of Charge}
\acrodef{soh}[SoH]{State of Health}
\acrodef{statcom}[STATCOM]{static synchronous compensator}
\acro{soec}[SOEC]{solid oxide electrolyzer cell}
\acro{tso}[TSO]{transmission system operator}
\end{acronym}

\maketitle

\begin{abstract}
The energy transition requires flexible technologies to maintain grid stability, and electrolyzers are playing an increasingly important role in meeting this need. While previous studies often question the dynamic capabilities of large-scale alkaline electrolyzer systems, we assess their potential to provide balancing services using real manufacturer data. Unlike common approaches, we propose the decoupling between the total electrolyzer power and a smaller fractions of power actually offered on balancing markets. Adapting an existing methodology, we analyze alkaline electrolyzer systems and extend the assessment to Germany and Europe. Our results show that large-scale electrolyzers are technically capable of delivering fast-response balancing services, with significantly lower dynamic requirements than previously assumed. The planned electrolyzers in Germany could cover the entire balancing capacity market, potentially saving around 13~\% of their electricity costs, excluding energy balancing revenues. The decoupling also resolves part of the trade-off for electrolyzer manufacturers, enabling the design of less dynamic but more stable systems. 
\end{abstract}


\section{Introduction}\label{sec:introduction}
\begin{figure}[htbp]
\centerline{\includegraphics[width=0.9\linewidth]{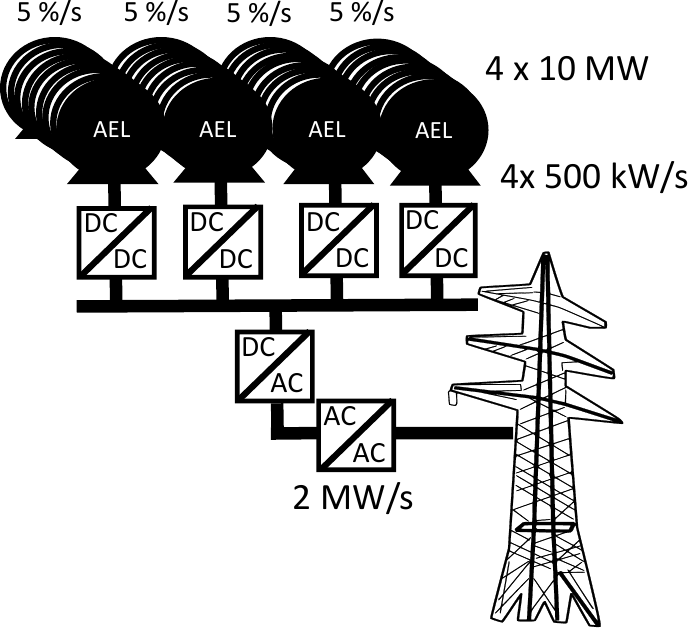}}
\caption{Schematic representation of four 10~MW alkaline electrolyzers with a ramp rate of 5\%/s, which together can offer 2~MW/s change to the grid.}
\label{fig:AEL_Schema}
\end{figure}

The global energy transition calls for scalable, low-carbon solutions. Hydrogen is increasingly seen as a cornerstone of future energy systems—particularly when produced via electrolysis using renewable electricity. The phase-out of conventional power plants with controllable generation and high rotational inertia is affecting grid stability and diminishing the availability of balancing services. Wind and solar power, while crucial for decarbonization, can currently provide these services only to a limited extent~\cite{wind_solar_ancillary}. Consequently, electrolyzers are increasingly being considered for grid support~\cite{samani2020grid}, particularly for the provision of \ac{fcr}, \ac{afrr} and \ac{mfrr}~\cite{cozzolino2024review}. Kopp et al. demonstrated in a real world setting at Energypark Mainz that electrolyzers can provide balancing capacity~\cite{EnergieParkMainz_PEM_aFRR}. Heynen et al.~\cite{heynen2024systematic} demonstrate in their review study the current state of research regarding the potential of large-scale electrolysis to reduce the rate of change of frequency and the maximum frequency deviation.
Studies repeatedly state that alkaline electrolysis is too slow to participate in \ac{afrr} or \ac{fcr} due to its insufficient dynamics~\cite{agredano2023dynamic}\cite{schmidt2017future}. 
In contrast, Eichman et al.~\cite{eichman2014novel} show, based on real systems, that changes in output can already be observed after just 0.2 seconds in both \ac{pem} and \ac{ael} systems. 
They have already demonstrated that both \ac{pem} and \ac{ael} are sufficiently dynamic in operation to provide grid services. 
Gu et al.~\cite{gu2024experimental} state that they require approximately 145 seconds for hot start-up in a 250 kW stack. This corresponds to a ramp rate of $\approx$ 0.68 \% / s.
They show that from a technical point of view, electrolyzers possess sufficient flexibility to participate in all of the aforementioned electricity markets. In a few seconds, these systems can adapt and provide power adjustments.
However, most conclusions are typically drawn from small-scale test stacks or simulation models. In such cases, the electrolyzer is often varied from its minimum to 100\% of its power, hence the full allowed range of the system. Cammann et al.~\cite{cammann2024dynamic} demonstrate in their study that \ac{fcr} through \ac{ael} can lead to cost reductions.\\
We examine large-scale electrolyzer systems using manufacturer data to assess their potential for participating in grid services. 
Unlike other approaches, we focus on systems that allocate only a small part of their capacity to such services. Studies and data indicate that the balancing capacity market is relatively small~\cite{regelleistung_dimensionierung}. 
The difference compared to other studies lies in the fact that, for large-scale electrolyzers, the dynamic response (in \%/s) refers to the system's total power, whereas balancing services are traded in fixed 1~MW increments~\cite{regelleistung2025}. We aim to raise awareness of this decoupling. Figure \ref{fig:AEL_Schema} illustrates why larger systems tend to exhibit a higher dynamic response per megawatt.

Figure \ref{fig:AEL_Schema} shows a configuration of four parallel alkaline electrolyzers connected to the high-voltage grid.
We apply our approach to large-scale electrolyzer~\cite{Dorn}, highlighting its limitations. In addition, we extend the analysis to Germany and Europe to examine how the comparatively low share of power dedicated to balancing services affects the overall potential. Our findings indicate that \ac{ael} can technically provide fast-response balancing services, thereby reducing the main operating cost — electricity — of electrolyzers by up to 13~\%. Our results show that only a small share of the planned electrolyzer power is needed to participate in the balancing market. This, in turn indicates that electrolyzers require significantly less dynamic operation than previously assumed. We also show that the decoupling helps manufacturers resolve the trade-off in electrolyzer design.
The findings provide a solid foundation for discussion.

The structure of the present paper is as follows: First, we provide an overview of the state of the art in Sec.~\ref{sec:stateoftheart}. Then Sec.~\ref{sec:method} outlines the methodology. In the evaluation sec. \ref{sec:Evaluation}, the approach is applied to different scenarios, including Germany and the EU. Finally, the Conclusion and Outlook in Sec. \ref{sec:Conclusion} summarize the key findings and suggest future works.

\section{State of the Art}\label{sec:stateoftheart}
This section outlines the relevant parameters, system characteristics, and physical relationships that form the foundation of the subsequent calculations. 
The section begins with an overview of the control power requirements, followed by a market analysis of exemplary manufacturers and their corresponding systems, including key technical parameters. Subsequently, the most relevant system limitations and boundary conditions are examined. Finally, the efficiency in the partial load range is described.

\subsection{Control power requirements}
Table \ref{tab:RegelleistungsArten} summarizes the requirements for the different types of balancing capacity. This includes the necessary ramp rates.
\begin{table}[htbp]
    \caption{Different balancing capacity mechanisms~\cite{Dorn}\cite{Regelleistung}.}
    \centering
    \begin{tabular}{|c|c|c|c|c|c|}
    \hline
       \textbf{Power} & \textbf{min. Size}   & \textbf{Availa.} & \textbf{Symmetry}& \textbf{Dur.}& \textbf{Gradient}\\
      \textbf{Control}   &    [MW]        & [sec.]              &  & [h]& [kW/s]\\
    \hline
           \textbf{\ac{fcr}} & 1 & 30 & sym. & 4 & 33.34 \\
    \hline
           \textbf{\ac{afrr}} & 1 & 300 & asym. & 4 & 3.34\\
    \hline
           \textbf{mFRR} & 1 & 750 & asym. & 4 & 1.34 \\
    \hline
    \end{tabular}
    \label{tab:RegelleistungsArten}
\end{table}
The table lists the type of balancing capacity and the minimum tradable power. It also specifies the availability, i.e., the latest point in time by which the sold power must be fully provided. Symmetry indicates whether the control reserve is offered symmetrically (in both directions) or separately for upward and downward regulation.
Durability describes the period for which the balancing capacity must be maintained. The gradient is calculated by dividing the provided power by the availability time.


\subsection{Market Availability of Industrial-Scale Electrolyzers}\label{sec:markedanalysis}
Selected market players are discussed to provide context for MW-scale electrolyzer dynamics: Key participants in the Chinese market include SANY Hydrogen Energy Co. Ltd.~\cite{SANY_E1200_2025} and Trina Green Hydrogen~\cite{trina2025one}, both of which currently offer \ac{ael} systems. In the American market, companies such as PlugPower Inc. and Cummins Inc. can be found, although not all dynamic performance data are available in their data sheets. In Europe, there is a wide range of companies offering \ac{ael}, \ac{aem}, \ac{soec}, and \ac{pem} electrolysis. A summary of the data on the dynamics can be found in Table \ref{tab:Elektrolyseure}, where necessary parameters can be obtained.

\begin{table}[htbp]
    \caption{Electrolyzer Market Dynamics Overview.}
    \centering
    \begin{tabular}{|c|c|c|c|c|}
    \hline
    \textbf{Name}                           & \textbf{Power}   & \textbf{Power Range}   & \textbf{Dynamic} & \textbf{Type}\\
    \textbf{}                               &    [MW]          &    [\%]                & [\% / sec.]      &                \\
    \hline
    \textbf{Ecolyzer~\cite{EcoLyzer}}       & 3                & 10-100                 & 0.5*              & \ac{ael}       \\
    \hline
    \textbf{Sunfire~\cite{SunfireHyLinkAEL2024}}       & 10                & 25-100                 & 0.61              & \ac{ael}       \\
    \hline
    \textbf{Sunfire~\cite{SunfireHyLinkSOEC2025}}       & 10                & 50-100                 & 0.16              & \ac{soec}       \\
    \hline
    \textbf{McPhy~\cite{McPhy}}       & 16               & 10-100                 & 5              & \ac{ael}       \\
    \hline
    \textbf{ThyssenKrupp~\cite{ThyssenKrupp}}       & 20               & 10-100                 & 3*              & \ac{ael}       \\
    \hline
    \textbf{Trina~\cite{trina2025one}}       & 15               & 30-110                 & 5              & \ac{ael}       \\
    \hline
    \textbf{QuestOne~\cite{questone_mhp_2024}}       & 10               & 10-100                 & 3*              & \ac{pem}       \\
    \hline
    \textbf{Elyzer~\cite{siemens_energy_electrolyzer_2024}}       & 17.5               & 40-100                 & 10              & \ac{pem}       \\
    \hline
    \textbf{Neptun ITM~\cite{itm_power_neptune_2024}}       & 2               & 25-100                 & 10              & \ac{pem}       \\
    \hline
    \textbf{Enapter~\cite{enapter_nexus2500_2025}}       & 2.5               & 1-100                 & 0.73              & \ac{aem}       \\
    \hline
    \multicolumn{4}{l}{\tiny{$^*$Discussions with manufacturers or calculated}}
    \end{tabular}
    \label{tab:Elektrolyseure}
\end{table}


\subsection{Participation of Electrolyzers in the balancing capacity Market}\label{sec:BalancingMarked}
In general, electrolyzers participate in the balancing capacity market in the same way as other units. For \ac{fcr}, it is important to offer symmetrical balancing capacity. For example, a product like NEG\_POS\_00\_04 covers a four-hour window between midnight and 4 a.m. The electrolyzer must therefore operate at a setpoint from which it can increase or decrease power output equally. For instance, a 100~MW plant operates at 95\% of its power and offers 5~MW as balancing capacity. The operator receives compensation for providing this 5~MW. Unlike with \ac{afrr}, there is no additional payment for actual activation—only for being available \cite{Regelleistung}.

\ac{fcr} responds automatically and decentrally based on the frequency profile. In contrast, \ac{afrr} responds to a control signal from the \ac{tso} and can be offered asymmetrically. An electrolyzer could, for example, run at 100 \% capacity and offer only downward regulation. In this case, the operator receives additional compensation, even if the power is never called upon. We do not consider \ac{mfrr} balancing market here, as it plays an increasingly minor role~\cite{Regelleistung}. To simplify the analysis, the balancing energy market is omitted here also, with the focus directed exclusively toward the balancing capacity market.

\subsection{Lower Power Limit}
The first crucial parameters for planning the operating strategy are the power limits \cite{Dorn}, expressed in percentages. The upper limit is typically set at 100~\%, even though some systems can operate at temporary overloads \cite{EcoLyzer}\cite{SunfireHyLinkAEL2024}\cite{ThyssenKrupp}. 
Various parameters in cell design are subject to trade-offs. Thinner membranes, for example, result in lower internal resistance and therefore reduced losses, but they also lead to increased hydrogen crossover~\cite{de2022optimal}. Additionally, factors such as cell area, current density, electrolyte, membrane material, pressure, and temperature can be optimized to favor one aspect over another~\cite{TradeOffCost_CurrentDensity}\cite{trinke2018hydrogen}\cite{kurzweil2018elektrochemische}\cite{TradeOff_Area_huang2024size}. The trade-off typically occurs between degradation (lifetime), costs, power, efficiency, and the dynamic behavior \cite{TradeOff_Channeldesign_PressureDrop_qi2023channel}.
In electrolysis systems, the \ac{hto} parameter is often specified. This indicates the proportion of hydrogen found in the oxygen stream. It plays a particularly important role during the minimum power of \ac{ael} systems~\cite{de2022optimal}. If the ratio becomes too high, explosive gas mixtures can form. This poses a safety risk and must therefore be avoided. Trinke et al. ~\cite{trinke2018hydrogen} investigate crossover phenomena in \ac{pem} and \ac{ael} electrolyis. At low production rates, the proportion of crossover becomes too large relative to the oxygen generation, leading to an increase in the \ac{hto}. For this reason, large-scale electrolyzers typically specify a minimum operating power between 10~\% and 50\% (Table \ref{tab:Elektrolyseure}).  

\subsection{Power Gradient}
The other important parameter for grid-friendly operation is the gradient, which describes how quickly the electrolyzer can ramp up or down. Some manufacturers differentiate between ramp-up and ramp-down, with ramp-down typically being significantly faster. Therefore, Table \ref{tab:Elektrolyseure} only shows the ramp-up values.
Electrochemistry in the stack reacts very quickly~\cite{eichman2014novel}. Large electrolyzers are more complex and require attention to more factors. High pressure helps the release of gases and allows higher ramp rates than at atmospheric pressure~\cite{RampRateEinschraenkung}. Temperature and pressure affect how electrolyzers work. Ogumerem and Pistikopoulos~\cite{RampRateEInschrankung2_Ogumerem} state that temperature mostly controls the dynamic behavior of the electrolyzer. Operators limit the ramp rate to keep the gas volume inside the cells within safe limits and to ensure proper gas venting~\cite{RampRateEinschraenkung}. The studies~\cite{RampRateEinschraenkung} and~\cite{KIAEE201540_X62} explain that the slowest variables—pressure, gas venting, and temperature—together with the control system, set the ramp rate limits. In discussions with \ac{ael} manufacturers (Ecoclean GmbH), it was noted that the dynamics of the stack are limited by the gas discharge.
A gas flow rate that is too high may damage the thin electrode plate, which must be strictly avoided. During ramp-down, less heat and gas are produced, allowing electrolyzers to shut down quickly. Thus the electrical part of the stack does not cause these limits.

\subsection{Partial Load Behavior of Electrolysis - Efficiency}\label{sec:Efficiency}
Several studies demonstrate the improved part-load behavior of electrolyzers within the approved operating ranges for common types of electrolyzers~\cite{WirkungsgradAEL}\cite{kiaee2015demonstration}\cite{KLOTZ201420844_wirkungsgradSOEC}.

Typical efficiency curves of electrolysis stacks initially show a steep increase up to a maximum point. After reaching this peak, they gradually decline in an almost linear fashion~\cite{EnergieParkMainz_PEM_aFRR}. The overall efficiency, or stack efficiency, is obtained by multiplying the Faradaic efficiency and the voltage efficiency~\cite{kurzweil2018elektrochemische}.
The low efficiency at low power levels results from the high share of activation polarization, which is represented in the models via the voltage efficiency term \cite{kurzweil2018elektrochemische}. Ohmic and diffusion losses play a less significant role in this operating range \cite{kurzweil2018elektrochemische}.
When operating within the datasheet-specified range~\cite{SunfireHyLinkAEL2024}, the power curve is approximately linear, and efficiency decreases linearly with increasing balancing capacity~\cite{WirkungsgradAEL}.




At higher current densities, the characteristic curve leaves the linear region, and diffusion polarization becomes dominant due to insufficient mass transport \cite{kurzweil2018elektrochemische}. This region is also avoided as an operating point, since the efficiency drops sharply here as well—similar to the region where activation polarization dominates.





     
    %


    
        %

\section{Methodology}\label{sec:method}
This section introduces the concept of decoupling. The previously defined parameters and constraints from the state of the art are then used to evaluate system performance within the method.

\subsection{Conceptual framework - The Decoupling}\label{sec:Framework}

The decoupling between the offered balancing capacity and the electrolyzer power becomes clear with a 100~MW electrolyzer that changes its power at a rate $\Delta_{el}$ of 1\% per second. It takes 40 seconds to ramp from 60 \% to 100 \% power (hot ramp-up). Therefore, offering 40~MW of symmetrical \ac{fcr} is not feasible (Table \ref{tab:RegelleistungsArten}~\cite{Regelleistung}). If only 5~MW of symmetric \ac{fcr} is provided, the required load change can be achieved within just 5 seconds—well below the 30-second response time threshold—making the system eligible for \ac{fcr}, with the same system.
The required ramp rate depends on the electrolyzer power. This concept applies only to large-scale electrolyzers, as small units do not provide enough power to join the power balancing market~\cite{Regelleistung}. The minimum size is calculated in~\cite{Dorn} and ranges between 1~MW and 4~MW, depending on the electrolyzer, its parameters and the type of balancing capacity provided. This also applies to large-scale electrolyzer installations consisting of several independent electrolyzer units, because for the grid, it makes no difference whether a 100~MW electrolyzer ramps up and down or ten 10~MW electrolyzers are controlled together.



\subsection{Calculation Method}\label{sec:calculation_method}
To determine size of an electrolyzer for participation in grid services 
Equation~\eqref{eq:DeltaEL} is used~\cite{Dorn}:

\begin{equation}\label{eq:DeltaEL}
    \frac{\Delta P}{\Delta t} = \frac{P_{el} \cdot (1 - u) \cdot \Delta_{el}}{\frac{P_{anc}}{P_{ts}}}
\end{equation}

where:
\begin{itemize}
    \item $P_{el}$ is the rated power of the electrolyzer,
    \item $u$ is the lower threshold of the electrolyzer in \%,
    \item $\Delta_{el}$ is the ramp rate of the electrolyzer (in \% per second),
    \item $P_{anc}$ is the amount of balancing capacity to be provided,
    \item $P_{ts}$ is the trading size, defined by grid code regulations.
\end{itemize}

Equation~\eqref{eq:DeltaEL} describes the power gradient that an electrolyzer must be able to achieve in order to deliver a certain amount of additional service power. The numerator reflects the technical capability of the electrolyzer, taking into account the rated power, the operational flexibility range, and the ramping behavior. The denominator accounts for market requirements by normalizing the desired balancing capacity $P_{anc}$ with the minimum tradable unit $P_{ts}$.

\section{Evaluation and Discussion}\label{sec:Evaluation}
In this section, we analyze a large-scale electrolyzer project to assess the technical limits of the decoupling. Then, we apply the decoupling concept to Germany and estimate the potential income through balancing services. We do this for both \ac{fcr} and \ac{afrr}. Finally, we take a brief look at the situation in Europe. Lastly, we discuss the advantages of decoupling in relation to the design trade-off faced by electrolyzer manufacturers.

\subsection{Demo4Grid Project}
In the EU project "Demo4Grid" a 4~MW Sunfire \ac{ael} was installed and operated in a grid-supportive manner~\cite{demo4grid2020}. Based on Table \ref{tab:Elektrolyseure}, it can be observed that the ramp rate is 0.61 \% per second and that the lower operating limit is 25 \% of the nominal power. For a 4~MW electrolyzer, this means that 3~MW (i.e., the range down to 25 \% load) can theoretically be made available for grid services. \ac{fcr} must be offered symmetrically, only half of this range—1.5~MW—can be considered. However, \ac{fcr} can only be offered in whole megawatts, which means that only 1~MW of \ac{fcr} can actually be provided.

This implies that the \ac{ael} would operate at 3~MW and be capable of providing \ac{fcr} in the range between 2~MW and 4~MW—assuming the required ramping can be achieved within 30 seconds. However, the Sunfire electrolyzer has a ramp rate of 0.61 \% per second. To reach the full 25 \% change in output, it would therefore require approximately 41 seconds, exceeding the \ac{fcr} requirement. As a result, this specific unit cannot participate in \ac{fcr}. Providing \ac{afrr} is feasible as can be seen in Table \ref{tab:RegelleistungsArten}.

By calculating backwards from the \ac{fcr} requirement, the minimum power of the electrolyzer can be derived: Given a ramp rate of 0.61 \%/s over 30 seconds, the allowed power range change would be limited to 18.3 \% of the power. Referring this to the required 25 \% load shift for 1~MW of \ac{fcr}, this would mean that the total installed power would need to be approximately 5.5~MW. In other words, if the electrolyzer had a nominal power of 5.5~MW, it would just be able to offer 1~MW of \ac{fcr} within the 30-second window.

This example calculation illustrates how to determine whether large-scale electrolyzers are eligible to participate in balancing services. Relatively small electrolyzer, such as the one used in the Demo4Grid project, quickly reach their dynamic limitations in this context. 

\subsection{German Use Case}
Between January 2022 and January 2024, approximately ±500~MW \ac{fcr} and ±2000~MW of \ac{afrr} was needed for Germany \cite{regelleistung_Daten}. 
Considering expansion targets with 10~GW electrolyzer power~\cite{bmwk_wasserstoffstrategie10GWElektrolyse}, it becomes evident that even partial provision of power from these systems should be sufficient to help maintain grid stability. If these 10~GW electrolyzers were to provide only 5\% (symmetrical: 90\%-100\% - standard load point 95\%) of their capacity for the \ac{fcr}, the 500~MW would already be covered.
Equation~\eqref{eq:DeltaEL} applied to 10~GW with a range of \(\pm\)~500~MW leads to the minimum gradient \(\Delta_{el,\text{min.}}\).\\
$P_{\text{anc}} = 2~\text{GW} ,\quad P_{\text{el}} = 10~\text{GW}, \quad P_{\text{ts}} = 2~\text{MW} \ (\text{symmetric})$
First the $\frac{\Delta P}{\Delta t}$ is calculated:
\[
  \quad \frac{\Delta P}{\Delta t} = \frac{1~\text{MW}}{30~\text{s}} = 34~\text{kW/s} = 0.034~\text{MW/s}
\]
The lower threshold \(u\) can be determined based on the required control energy. With 10~GW of installed capacity and a symmetric need of 500~MW, the usable range is 1~GW. This results in:
\[
10~\text{GW} - 1~\text{GW} = 9~\text{GW} \Rightarrow u = \frac{9~\text{GW}}{10~\text{GW}} = 0.9
\]

After substituting all values and rearranging the equation for \(\Delta_{el}\), the gradient is obtained.

\[
\Delta_{el} = 0.0945\%/\text{s}
\]

This means that if the average ramp rate of the electrolyzers in Germany reaches 0.0945~\%/s, they can fully provide the currently required \ac{fcr}. All electrolyzers listed in Table \ref{tab:Elektrolyseure} can handle this, at least in theory, even the \ac{soec}, although it probably does not initially serve this purpose.
If the remaining 9~GW is used for the $\pm$2GW \ac{afrr} with u=0.56 and the minimum gradient for \ac{afrr} (0.0034~MW/s), a $\Delta_{el.}$ of 0.086~\%/s results. Electrolyzers can also handle this. In summary, the planned expansion of the electrolyzer capacity in Germany by 2030 allows electrolyzers to provide the entire balancing capacity for \ac{fcr} and \ac{afrr} as of today. This shows a decoupling between the balancing capacity an electrolyzer provides and its full power, which allows the use of any electrolysis technology.

\subsection{Example calculation Frequency Containment Reserve}
To approximate additional revenue using \ac{fcr} for electrolysis, data from the \ac{tso} is used~\cite{regelleistung_Daten}. Between July 16 and July 22, 2025, prices range from 15 €/MW at night to 120 €/MW at midday. The weekly sum is 2,350 €/MW/week. This so-called capacity price is publicly available for previous years. In 2025, capacity prices range between 1,300 and 5,877 €/MW/week. 
In the future, more providers are expected to enter the market. This trend could push prices down. Rapid expansion of large-scale batteries, virtual power plants, heat pumps, and electrolyzer capacity drives this development~\cite{NextKraftwert}. For now, we base our calculation on the prices from July 17, 2025, as shown in Table \ref{tab:FCRPrices_17_07_2025}, which represents a typical weekday and is used here for illustrative purposes only.
\begin{table}[htbp]
    \caption{FCR capacity prices for different time blocks (17.07.2025).}
    \centering
    \begin{tabular}{|c|c|}
    \hline
    \textbf{Time Block} & \textbf{Price [€/MW]} \\
    \hline
    NEGPOS\_00\_04 & 14.71 \\
    \hline
    NEGPOS\_04\_08 & 21.92 \\
    \hline
    NEGPOS\_08\_12 & 62.00 \\
    \hline
    NEGPOS\_12\_16 & 78.00 \\
    \hline
    NEGPOS\_16\_20 & 51.00 \\
    \hline
    NEGPOS\_20\_24 & 36.00 \\
    \hline
    \end{tabular}
    \label{tab:FCRPrices_17_07_2025}
\end{table}
An electrolyzer with a capacity of 100~MW, operating continuously at 95 \% load and offering 5~MW of \ac{fcr}, could have earned an additional 1,318.15 € in \ac{fcr} compensation on July 17, 2025, based on the applicable time-block prices. According to~\cite{Dorn}, the overnight \ac{capex} for an \ac{ael} is between 40–83 million € for a 100~MW electrolyzer. However, the primary cost factor in electrolysis is electricity~\cite{felgenhauer2015state}, which is why we compare the revenues from balancing services directly to electricity expenses. Using German electricity prices from 2024~\cite{smard}, and considering only the hours when the wholesale electricity price is below 85 €/MWh—in order to ensure more than 5,000 operating hours per year—we arrive at an average market price of 50 €/MWh.
Electrolyzers are exempt from electricity tax~\cite{StromsteuerEntfaellt}. They are also exempt from grid fees only if the electricity is later fed back into the electricity grid~\cite{NetzEntgelteSpeicher}. We assume the hydrogen produced will be used in the chemical industry and not re-injected, we include grid charges in our estimate. Assuming grid fees at 30 \%, an electricity price of 65 €/MWh results.
For a set point of 95 MW running 24 hours per day, corresponds to electricity price of:
\[
   95~\text{MW} \cdot 24~\text{h} \cdot 65~\text{€/MWh} = 150,000~\text{€}
\]
Providing \ac{fcr} can reduce these electricity costs by around 1~\%, and additionally, \ac{fcr} NEG effectively provides free energy to the electrolyzer during activation periods.

\subsection{Example calculation  automatic Frequency Restoration Reserve}
For an electrolyzer, it usually makes more sense to offer positive \ac{afrr}, since the unit can operate at 100 \% power and only needs to reduce load when instructed. If no activation occurs, the operator still receives the basic capacity marked payment without delivering balancing energy. 
The capacity price for awarded \ac{afrr} contracts typically ranges from a few euro cents up to around 100 €/MW per hour. For estimation purposes, we assume an average of 20~€/MW/h. One key advantage of providing positive \ac{afrr} is that electrolyzers are typically capable of executing a rapid hot ramp-down reliably and quickly at any time. This allows them to offer flexibility across their full operational range. 

%

Assuming 5~MW (symmetrical) of a 100~MW \ac{ael} are reserved for \ac{fcr}, and a minimum load of 50~\% is allowed, then up to 40~MW remain available for \ac{fcr}. For a full day, this results in the following revenue:
\[
   40~\text{MW} \cdot 24~\text{h} \cdot 20~\text{€/MW} = 19,200~\text{€/day}
\]
This corresponds to approximately 12~\% of the electricity costs calculated in the previous section. In addition, the provider receives compensation for the actual activated energy, which can significantly increase total income. The activation price (i.e. the price for actually delivered energy) for \ac{afrr} POS often varies between 30 and 15,000 €/MWh, the latter being a predefined price cap. These prices are highly dependent on time of day and market conditions, so the values mentioned here should only be seen as approximate benchmarks. For a calculation, an average activation price of 1,000 €/MWh in addition to the capacity price could be applied in a more in-depth work.
However, this data is anonymized and therefore not included in the calculation.\\
The \ac{afrr} participation also comes with a trade-off: Electrolyzers are primarily intended to produce hydrogen. Frequent participation in \ac{afrr} POS leads to reduced full-load hours per year, lowering overall hydrogen output.

\subsection{European Use Case}
The Central European \ac{fcr} Cooperation currently requires 1500~MW of \ac{fcr}~\cite{Regelleistung}, while the Nordics and the rest of Europe need an additional 1500~MW. This results in a total demand of 3000~MW of balancing capacity. The EU plans to utilize 80 GW of electrolysis capacity by 2030, with 40 GW to be installed within the EU and another 40 GW outside its borders~\cite{EHO2025}. This implies that only about 15 \% (7.5 \% symmetrical) of the planned electrolysis capacity would be required to provide the entire \ac{fcr}. The ratio between required balancing capacity and Europe’s electrolyzer expansion targets is lower, but a significant share can still be provided this way.

\subsection{Trade-off in electrolyzer design}
\begin{figure}[htbp]
\centerline{\includegraphics[width=0.9\linewidth]{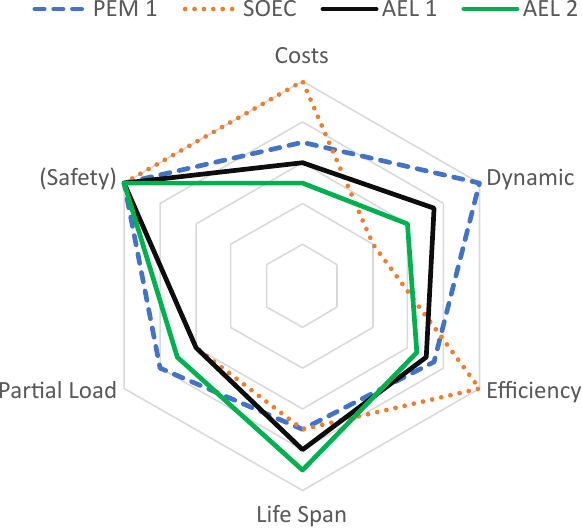}}
\caption{Schematic representation of the trade-off in electrolyzer design.}
\label{fig:TradeOff}
\end{figure}

The design of electrolyzers is a complex topic. Entire studies and academic works are dedicated to it, as can be seen in section \ref{sec:method} \cite{TradeOffCost_CurrentDensity}\cite{TradeOff_Area_huang2024size}\cite{TradeOff_Channeldesign_PressureDrop_qi2023channel}. Therefore, this work aims only to provide a brief overview, of the key aspects. Figure \ref{fig:TradeOff} tries to illustrate the different design objectives that can guide the development of an electrolyzer. 

These objectives help explain why the electrolyzers listed in Table \ref{tab:Elektrolyseure} differ so significantly in their technical parameters.
One example is the membrane thickness: choosing a thicker membrane increases internal resistance but reduces gas crossover \cite{de2022optimal}\cite{trinke2018hydrogen}. While this may negatively impact efficiency, it can enable operation at lower partial loads, allow for higher operating pressures (which can improve BoP efficiency), and increase system safety. However, these advantages may come at the cost of higher system complexity and expense. Given that safety is a parameter that must be met, it is indicated in parentheses in the figure. 

Similarly, increasing system dynamics may lead to more gas bubble formation within the stack, which in turn requires additional measures to remove these bubbles and maintain low diffusion polarization—again resulting in higher costs \cite{TradeOffCost_CurrentDensity}\cite{kurzweil2018elektrochemische}. Such trade-offs influence many of the degrees of freedom in electrolyzer design, and are the reason for the wide variation among different systems. Additional influential parameters include electrode thickness, flow field design (channel geometry), electrolyte composition, and others~\cite{de2022optimal}\cite{trinke2018hydrogen}\cite{TradeOff_Channeldesign_PressureDrop_qi2023channel}\cite{TradeOffszhang2020electrochemical}.

The decoupling between the offered control power and the actual electrolyzer power, as discussed in this work, enables manufacturers to eliminate certain design constraints and optimize for other advantages. For instance, if there is no longer a need to operate at very low partial loads—because the system only runs between 90~\% and 100~\% of nominal capacity—the membrane can be made thinner, potentially increasing overall efficiency~\cite{TradeOffszhang2020electrochemical}. Additionally, the required ramp rate (gradient) for large-scale systems becomes significantly less demanding, allowing systems with lower gradients to still participate in \ac{fcr} markets.

\subsection{Competing balancing capacity participants}
The list of awarded bids for \ac{fcr} is anonymized, which makes it difficult to trace the underlying technologies~\cite{regelleistung_Daten}. Based on the regulatory framework, it can be inferred that primarily conventional power plants participate in \ac{fcr} balancing~\cite{Regelleistung}. Since a guaranteed duration of 4 hours (Table \ref{tab:RegelleistungsArten} is required, not every battery system can participate; large-scale Batteries are particularly well suited. These are generally limited by the amount of energy they can absorb or deliver. Additionally, pumped-storage power plants, hydropower plants, biogas plants, and waste-to-energy facilities can participate in the tenders. Large industrial plants are also well suited for load shedding and load increase. Increasingly, virtual power plants and large heat pumps are entering this market~\cite{NextKraftwert}. As conventional power plants are gradually phased out from continuous operation, there is a growing need for new actors to fill this gap. Electrolyzers can increasingly play an important role for this, even though they will face competition from many existing market participants. In times when renewable energy is scarce, electrolyzers may, therefore, be completely shut down and cannot support the grid stability. Gas-fired power plants (capable of hydrogen combustion) are currently planned for these times.
\section{Conclusions and Outlook}\label{sec:Conclusion}
In this work, we discuss the use of electrolyzers for providing Frequency Containment Reserve (FCR) and automatic Frequency Restoration Reserve (aFRR).
We show that Alkaline Electrolyzers (AEL) can operate dynamically to deliver balancing capacity in a conventional way. The key point is the decoupling between the power of the electrolyzer and the offered balancing capacity. This implies that most systems can provide balancing capacity by reserving only a small part of their upper power range.
The key parameters for the calculations are the hot ramp-up gradient, the minimum partial load, and the nominal power of the electrolyzer. These parameters involve a trade-off in electrolyzer design, which can be mitigated by decoupling the offered flexibility from the installed nominal capacity. This approach can be used to improve both lifetime and efficiency and to be sufficiently dynamic.
We discuss the limits of decoupling: In particular, we show that \ac{ael} can participate in balancing capacity marked.
The German government's target for electrolyzer expansion is compared with the total balancing capacity required. Because the need for balancing capacity is relatively low, electrolyzers could completely cover the balancing capacity need. This could lead to lower prices for \ac{fcr} and \ac{afrr} as the number of electrolyzers increases.
In the EU, 15~\% of the electrolysis capacity would be sufficient to cover \ac{fcr}.
The main cost component of electrolysis is the electricity requirement. The study shows that participation in the balancing capacity marked could save 13~\% of electricity costs, when the revenues from \ac{fcr} and \ac{afrr} are combined.\\     

Electrolyzers can react quickly to changes and will probably expand significantly across Europe. The decoupling we describe in this work also applies to inertia provision. Other relevant research areas for large-scale electrolyzers include voltage regulation, reactive power support, and grid-forming load behavior.





    


\printbibliography 

\end{document}